\documentclass[apj]{emulateapj}

\usepackage{amsmath}
\usepackage[normalem]{ulem}

\newcommand{\be}{\begin{equation}}
\newcommand{\ee}{\end{equation}}
\def\la{\buildrel < \over {_{\sim}}}
\def\ga{\buildrel > \over {_{\sim}}}

\shorttitle{Neutral return flux and its effects on acceleration of test particles}
\shortauthors{Blasi et al.}

\begin{document}

\title{Collisionless shocks in a partially ionized medium:\\
I. Neutral return flux and its effects on acceleration of test particles}

\author{P. Blasi\altaffilmark{1}, G. Morlino\altaffilmark{1}, R.
Bandiera\altaffilmark{1}, E. Amato\altaffilmark{1}}
\affil{INAF-Osservatorio Astrofisico di Arcetri, Largo E. Fermi, 5, 50125
Firenze, Italy}

\and

\author{D. Caprioli\altaffilmark{2}}
\affil{Department of Astrophysical Sciences, Peyton Hall, Princeton University,
Princeton, NJ 08540, USA}

\begin{abstract}
A collisionless shock may be strongly modified by the presence of neutral atoms
through the processes of charge exchange between ions and neutrals and
ionization of the latter. These two processes lead to exchange of energy and
momentum between charged and neutral particles both upstream and downstream of
the shock. In particular, neutrals that suffer a charge exchange downstream with
shock-heated ions generate high velocity neutrals that have a finite probability
of returning upstream. These neutrals might then deposit heat in the upstream
plasma through ionization and charge exchange, thereby reducing the fluid Mach
number. A consequence of this phenomenon, that we refer to as {\it the neutral
return flux,} is a reduction of the shock compression factor and the formation
of a shock precursor upstream. The scale length of the precursor is determined
by the ionization and charge exchange interaction lengths of fast neutrals
moving towards upstream infinity. 
In the case of a shock propagating in the interstellar medium, the effects of
ion-neutral interactions are especially important for shock velocities
$<3000$~km~s$^{-1}$.
Such propagation velocities are common among shocks associated with
supernova remnants, the primary candidate sources for the acceleration of
Galactic cosmic rays. We then investigate the effects of the return flux of
neutrals on the spectrum of test-particles accelerated at the shock. We find
that, for shocks slower than $\sim 3000$~km~s$^{-1}$, the particle energy
spectrum steepens appreciably with respect to the naive expectation for a strong
shock, namely $\propto E^{-2}$.
\end{abstract}

\keywords{ acceleration of particles -- atomic processes -- line:profiles --
ISM: supernova remnants}

\section{Introduction}
 \label{sec:intro}

The Physics of collisionless shock fronts is of paramount importance for
numerous astrophysical sources, from supernova remnants (SNRs) and shocks in the
Solar System to gamma-ray bursts and active galactic nuclei, as well as for the
description of the process of cosmic ray acceleration. Such shocks are formed
due to the mediation of electromagnetic instabilities while particle collisions
are insignificant, thereby the name of {\it collisionless}. The shock itself,
or the subshock in the case of a cosmic-ray (CR) modified shock,
forms on a spatial scale of the order of $\sim \xi r_{L,th}$, where $\xi>1$ and
$r_{L,th}\approx 10^{10} B_{\mu} T_{8}^{1/2}$ cm is the gyration radius of
thermal particles at a temperature $T=10^{8}T_{8}~K$ and in a background
magnetic field $B_{\mu}~\mu$G \footnote{Rigorously this statement applies to
quasi-perpendicular shocks, while parallel shocks require a more detailed
discussion. However for particle accelerating shocks, substantial magnetic field
amplification must take place upstream, so that even if the shock configuration
is initially parallel, due to compression, the magnetic field in the downstream
plasma becomes mostly oblique.}. The shock formation is associated with the
development of electromagnetic streaming instabilities ({\it e.g.} Weibel
instability),  which can amplify or even create magnetic fields \cite[]{weibel}.
Interestingly, the formation of collisionless shocks has recently started being studied
in the laboratory by creating a laser-driven plasma expansion 
\cite[]{lab1,lab2,lab3,lab4}.

By definition, collisionless shocks may only develop in ionized media. On the
other hand, most astrophysical plasmas contain some fraction of neutral
material, which will behave differently from the ionized component at the
crossing of the shock.

Ions are basically isotropized at the shock and heated to a temperature that
mirrors the ram pressure of the incoming ionized fluid. If electrons are subject
to the same fate, their temperature immediately behind the shock is
$m_{e}/m_{p}$ times the temperature of protons (ions). The temperature of
electrons and protons may in fact turn out to be closer to each other due to the
electromagnetic coupling between the two components \citep[see e.g.][and
references therein]{cp88,G+07}. In addition, Coulomb scattering will also act
towards temperature equilibration. This last process however, typically acts on
much longer time-scales, which in the case of several observed SNRs  \citep[see
e.g.][]{ellison10}, actually exceed the age of the source.

As to the neutrals, these are insensitive to the shock transition and, to
zeroth order approximation, they can cross the shock surface without interacting
with the ions. The interaction of neutrals and ions occurs mainly through charge
exchange and ionization and can change the structure of the shock rather
dramatically because of the energy and momentum deposition that is involved.

The cross section of these processes is of order $\sim 10^{-15}~\rm cm^{2}$,
therefore, even in relatively tenuous plasmas, the rate at which they occur can
lead to phenomena of potential astrophysical importance. 
Both charge exchange and ionization rates are proportional to the relative
velocity between neutrals and ions.

Upstream of a classical shock one expects that neutrals and ions are in some
sort of local thermal equilibrium (as in the unshocked ISM), therefore they will
have the same velocity and the same temperature. In these conditions there is
neither net ionization, nor net charge exchange: clearly charge exchange will
occur anyway because of the different velocities of neutrals and ions in
different parts of their respective thermal distributions, but without changing
the particle distributions. On the other hand, the neutrals crossing the shock
front experience a thermal bath of hot ions with a bulk motion which is slower
than their own ($\sim 4$ times slower for a strong shock, if not modified by
CRs). Moreover, while neutrals remain at their upstream temperature, ions are
heated in the shock transition. As a result, a net velocity difference arises
between the two components and both ionization and charge exchange {\it are
turned on}, namely become effective at modifying the distribution functions of
both species.

Therefore, when a shock propagates into a partially ionized medium,
an interesting chain of processes develops. When fast, cold
neutrals undergo charge-exchange interactions with the slow hot ions downstream
of the shock, some fraction of the resulting hot neutrals can cross the shock
and move towards upstream infinity: a "return flux" develops. The relative
velocity between these hot neutrals and the upstream ions triggers the onset of
charge-exchange interactions that lead to the heating and slowing down of the
ionized component of the upstream fluid. The system tends to develop a shock
precursor, in which the fluid velocity gradually decreases from its value at
upstream infinity. As soon as the ions develop a velocity gradient in the
upstream, charge exchange interactions become effective at modifying the
particle distribution functions here as well, in a complex non-linear chain, in
which the information is carried from the downstream to the upstream by the
return flux of neutrals. The consequent reduction of the shock Mach number has
potential implications, in turn, on the physics of dissipation and particle
acceleration.

The presence of neutrals results in a shock modification that is qualitatively similar, in some 
respects, to the one induced by CR acceleration, for instance in SNRs, where the 
efficiency of acceleration is expected to be between 5 and 15\%, according to the most recent 
estimates of \cite{ba12}. The shock dynamics is profoundly changed even with such modest acceleration
efficiency: the dynamical reaction of accelerated particles lead to the formation of a precursor
upstream of the shock (see e.g. \cite{reviewNLA} for a review) and to magnetic field amplification
due to streaming instability, a crucial ingredient to accelerate particles to very high energies. 
The velocity compression factor felt by particles is a function of momentum, which results in
concave spectra steeper at low energies than they are at high energies (e.g. \cite{ab05,be99}). 
Moreover, the particles' escape from upstream infinity at the maximum momentum allows for 
the total compression factor to become $>4$ for strong shocks, thereby leading to spectra harder 
than $E^{-2}$ at energies larger than a few GeV. 

In both cases of a CR induced or neutrals induced precursor, the subshock is weakened and 
steeper spectra of accelerated particles can be expected. However, when looked up in detail,
the two cases are very different: first, for CR induced modification, the fraction of energy 
deposited in the form of thermal energy of ions upstream is much smaller than for the case 
of modification induced by neutrals, at least for shocks with velocity smaller than $\sim 3000$
km/s. Moreover, while the typical scale-length of a CR induced precursor is set by the diffusion 
length of the particles that carry most energy (this in turn depends on the spectrum of accelerated 
particles), the scale-length of the precursor induced by neutrals is set by the charge exchange and
ionization cross sections and by the ion and neutral densities. While the CR induced precursor 
always leads to spectral steepening at energies below $\sim 10$ GeV, in the case of the precursor 
induced by the neutral return flux, the subshock weakening can easily reflect in spectral steepening
up to much larger energies. While the interaction between the two shock modifications can lead to 
interesting phenomena, here we limit ourselves with solving the difficult problem of deriving the structure 
of a collisionless shock propagating in a partially ionized medium, when CRs can be treated as test particles.
The most complete problem in which the CR dynamical reaction is taken into account
will be discussed in a forthcoming paper.

The problem of describing a shock transition occurring in a partially neutral
fluid can be treated in different ways but the general solution is known to
consist of a final state (downstream infinity) in which neutrals eventually
disappear through ionization and the ion density increases. This final state can
be determined by adopting a fluid approach between upstream infinity and
downstream infinity and simply assuming conservation of the fluxes of mass,
momentum and energy \cite[]{Morlino10}. 
It is however not possible to use a naive two (or multi) fluid approach to
adequately describe the complex system of neutrals and ions. This is because
neutrals can only be taken to behave as a fluid on scales that are usually much
larger than those of phenomenological importance. On such large scales, the
system behaves as if the shock had become collisional (the collisions being
represented by charge exchange interactions): also the neutrals are then
"shocked", through a transition region that has now become much thicker. At the
same time, on a comparable length-scale, the neutral component gradually
disappears because of ionization.

Studies on this subject traditionally focused on the solar wind termination
shock, but it is easy to see how the problem is extremely relevant also in the
context of supernova remnant shocks. These shocks sometimes propagate in the
cold, weakly ionized ISM, as shown by the associated $H\alpha$ emission. In
addition, several remnants showing an interesting phenomenology are actually
thought to be interacting with dense cold clouds of quasi-neutral material. 

The $H\alpha$ emission provides a powerful diagnostic tool for the
conditions at the shock. As first recognized by \cite{Chev-Ray78} and
observed by \cite{Chevalier80}, the $H_\alpha$ profile detected in association with SNR shocks usually
consists of two components, a narrow one, whose width reflects the temperature
of the upstream medium, and a broad one, due to neutrals that have undergone
charge-exchange with the hot downstream protons. This second component
represents a unique tool to {\it measure} the temperature of ions, usually very
difficult to access otherwise. After the pioneering work of \cite{Chev-Ray78}
and \cite{Chevalier80} several authors have further refined the use of Balmer
emission as a diagnostics for SNR shocks.
Among others, \cite{Ghavamian01} incorporated Monte Carlo modeling of Lyman
$\alpha$ absorption, while \cite{HengA07} and \cite{HengB07} included a more
careful evaluation of the charge-exchange cross section and
\cite{vanAdelsberg08} considered the effect of multiple charge exchanges on
the distribution function of the hot neutrals. However, all these works present
limitations in the calculation of the distribution function of neutrals and do
not take into account the effect of the return flux.

In the solar environment these processes are also important, but the spatial scales involved 
are much smaller and the expected phenomena have different manifestations. For instance \cite{zank96} 
discussed a similar effect of returning neutrals on the heating of the heliopause, as inferred from 
Ly-$\alpha$ absorption. But the ideal test bench for the phenomenon of neutral return flux is
actually represented by SNRs, given that for typical densities of the ISM the diffusion length of
energetic particles upstream is in this case comparable with or in excess of the interaction length for
charge exchange and ionization.

The final goal of our investigation, to be illustrated in a forthcoming paper,
is to use Balmer lines as a diagnostic tool for the detection of efficient
CR acceleration in SNRs: if particle acceleration is effective, a
sizeable fraction of the ram pressure is transformed into non-thermal particles,
thereby reducing the heating at the shock surface. This should reflect in a
lower ion temperature downstream of the shock, and hence in a narrower broad
Balmer line. On the other hand, efficient particle acceleration also leads to
the formation of a precursor upstream of the shock, where charge exchange can
then occur. This leads to ion heating upstream, and correspondingly to a broader
narrow Balmer line. \cite{Wagner09} were the first to incorporate ionization,
emission, and heating in a CR precursor into the modeling of the H$\alpha$
emission treating both CRs and neutrals as fluids. A similar effort has be done
by \cite{Raymond11} assuming a parametric structure for the CR precursor.
Interestingly, several signatures of phenomena induced by the CR acceleration
have been collected in the last few years \cite[see e.g][for a review]{Heng09},
but the absence of an appropriate mathematic tool still prevents us from describing
them in a physically satisfactory way. Providing such a tool is the purpose of
our series of papers, the present one being the first, while other implementations,
like the inclusion of dynamically relevant CRs, will be the scope of forthcoming papers.
 
The paper is organized as follows: in \S~\ref{sec:math} we write down the
mathematical formalism for the solution of Vlasov equation for the neutrals and
of the fluid equations for the ions. In \S~\ref{sec:results} we illustrate the
main results in terms of modified shock dynamics, test particle acceleration in
the presence of neutrals and Balmer emission. We conclude in
\S~\ref{sec:conclude}.

\section{The mathematical approach}
\label{sec:math}

\begin{figure}
\begin{center}
\includegraphics[width=0.47\textwidth]{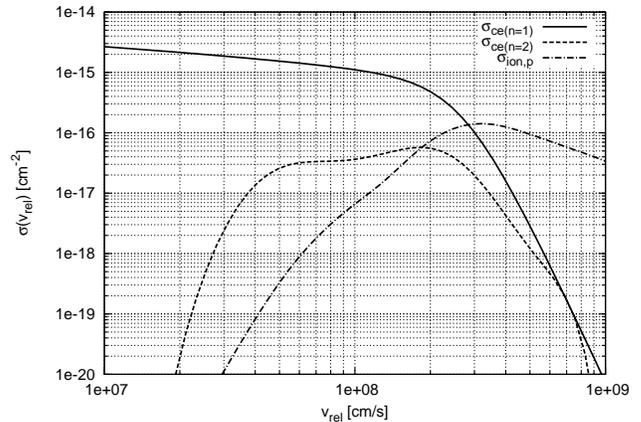}
\end{center}
\caption{Cross section for charge exchange with $n=1$ hydrogen atoms (solid
line)  and with $n=2$ hydrogen atoms (dashed line) {\protect \cite[]{Barnet90}}.
The dotted
line is the cross section for ionization of neutral Hydrogen due to collisions
with protons {\protect \cite[]{Janev93}}.
All cross sections used are provided by the International Atomic Energy Agency
via the web-site \texttt{http://www-amdis.iaea.org/ALADDIN/}.}
\label{fig:cross}
\end{figure}

The distribution function $f_{N}(\vec v,\vec x)$ of neutrals interacting with
ions through charge exchange and ionization is described by the Vlasov equation
\be
\frac{\partial f_{N}}{\partial t} + \vec v \cdot \nabla f_{N} = \beta_{N} f_{i} 
- \beta_{i} f_{N},
\label{eq:vlasov}
\ee
where $f_{i}$ is the distribution function of ions, assumed here to be a
drifting Maxwellian distribution at a temperature $T$ dependent upon location,
with a bulk velocity representing the local speed of the ion plasma (see
Eq.~\ref{eq:max} below). In Eq.~\ref{eq:vlasov} we have introduced
\be
\beta_{i} (\vec v) = \int d^{3} w\ v_{rel}\ \left[\sigma_{ce}(v_{rel}) +
\sigma_{ion}(v_{rel})\right]\ f_{i}(\vec w),
\ee
\be
\beta_{N} (\vec v) = \int d^{3} w\ v_{rel} \ \sigma_{ce}(v_{rel})\ f_{N}(\vec
w),
\ee
and $v_{rel}=|\vec v - \vec w|$. The quantity $\beta_{i}$ represents the rate of charge
exchange (cross section $\sigma_{ce}$) and ionization (cross section
$\sigma_{ion}$) of a neutral with velocity $\vec v$, while $\beta_{N}$ is the
rate of charge exchange of an ion that becomes a neutral. The cross sections are
all functions of the modulus of the relative velocity between an ion and a
neutral and are plotted in Fig.~\ref{fig:cross}. The ionization cross section is
shown as a dash-dotted line. $\sigma_{ce}$ is the charge exchange cross section
for hydrogen atoms including all possible final states. In Fig.~\ref{fig:cross}
the solid line refers to the cross section when the final state is the ground
state, $n=1$, while the dashed line refers to charge exchange with final
state $n=2$ \cite[]{Barnet90}. Following \cite{HengA07}, in order to
obtain cross sections for $n > 2 $ we use the scaling relation $\sigma_{ce,n} =
(2/n)^3\, \sigma_{ce,2}$, first proposed by \cite{Janev93}. Hence the total
charge exchange cross section can be written as:
\begin{equation} \label{eq:sigma_ce}
 \sigma_{ce} = \sum_{n=1}^{\infty} \sigma_{ce,n}
 \approx \sigma_{ce,1} + \sigma_{ce,2} \sum_{n=2}^{\infty} \left( 2/n
 \right)^3 \,.
\end{equation}

For $\sigma_{ion}$ we consider here only the ionization due to collisions with
protons. For the sake of simplicity, we neglect the contributions to ionization
due to collisions with electrons or heavy ions. These contributions may actually
be comparable to that of protons, and therefore relevant for the calculation of
the total Balmer emission \cite[see e.g. Figure 1 of][]{HengA07}. We will
introduce them in future works.

In solving Eq.~\ref{eq:vlasov} we assume that a stationary situation is
reached, so that $\partial f_{N}/\partial t=0$. Moreover we restrict our
attention to the case of a plasma moving in one direction, $z$, so that the
problem reduces to describing the evolution in the directions parallel
($\parallel$) and perpendicular ($\perp$) to $z$. Eq.~\ref{eq:vlasov} is then
reduced to:
\be
v_{\parallel} \frac{\partial f_{N}}{\partial z} = \beta_{N} f_{i} - \beta_{i}
f_{N},
\ee
where we used the fact that there cannot be gradients of the distribution
function in the directions perpendicular to the symmetry axis of the problem,
$z$. 

We describe here a novel way to solve the Vlasov equation, based on decomposing
the distribution function $f_{N}$ in the sum of the neutrals that have suffered
0, 1, 2, ... , $k$ processes of charge exchange. Each distribution function is
named $f_{N}^{(k)}$, and clearly $f_{N}=\sum_{k}f_{N}^{(k)}$. 
Since the neutrals do not feel  the shock directly, for them there is no
distinction between upstream and downstream except for the interaction with
different populations of ions on the two sides of the shock front. Upstream ions
and neutrals are assumed to start (at $z\to -\infty$) with the same bulk
velocity and temperature, therefore charge exchange occurs at equilibrium and
the distributions do not change appreciably (but see discussion below). On the
other hand, after entering the downstream, the neutrals experience a very
different environment and both ionization and charge exchange occur effectively.

Let us consider the equation describing $f_{N}^{(0)}$, namely the distribution
of particles that have not suffered any charge exchange. This has no source
terms and is easily seen to be:
\be
v_{\parallel} \frac{\partial f_{N}^{(0)}}{\partial z} = - \beta_{i} f_{N}^{(0)},
\ee
with solution:
\be
f_{N}^{(0)} (z,v_{\parallel},v_{\perp}) = f_{N}^{(0)} (-\infty,
v_{\parallel}, v_{\perp}) \, e^{-\int_{-\infty}^{z} \frac{d
z'}{v_{\parallel}} \beta_{i} (z',v_{\parallel},v_{\perp})} \,.
\ee
Quite obviously, the number density of neutrals that did not suffer any charge
exchange decreases exponentially after one interaction length, as defined by the
ratio between $v_\parallel$ and the rate $\beta_{i}$. It is worth noting that, in
a plasma at equilibrium, charge exchange does occur although the net number of
neutrals in a given volume of phase space does not change. This means that the
position $z=-\infty$ is somewhat arbitrary and simply needs to be chosen far
enough from the shock surface. It also means that in order to obtain the actual
solution of the problem, one has to sum up a sufficiently large number of
$f_{N}^{(k)}$. 

In the absence of any net velocity difference between ions and neutrals the sum
of all $f_{N}^{(k)}$ must return identically $f_{N}^{(0)}(z= -\infty,
v_{\parallel}, v_{\perp})$ for any $z$. The equation describing any
$f_{N}^{(k)}$ is:
\be
v_{\parallel} \frac{\partial f_{N}^{(k)}}{\partial z} = \beta_{N}^{(k-1)} f_{i} 
- \beta_{i} f_{N}^{(k)}.
\label{eq:vla}
\ee
The formal solution of this equation, as derived through elementary methods is:
\begin{eqnarray}
f_{N}^{(k)}(z) = \int_{-\infty}^{z} \frac{dz'}{v_{\parallel}}
\beta_{N}^{(k-1)}(z') f_{i} \exp \left[ \int_{z}^{z'} \frac{dz''}{v_{\parallel}}
\beta_{i}(z'') \right]  \nonumber \\
+ \lambda \exp\left[ -\int_{-\infty}^{z} \frac{dz''}{v_{\parallel}}
\beta_{i}(z'') \right],
\end{eqnarray}
where the integration constant $\lambda$ has to be determined by using the boundary
conditions of the problem. These are different for $v_{\parallel}>0$ and
$v_{\parallel}<0$. Let us first consider the case $v_{\parallel}<0$: for $z\to
+\infty$ (downstream infinity) the number of particles that have suffered $k$
scatterings must vanish, therefore, since $-\int_{-\infty}^{\infty} dz/v_{\parallel}
\beta_{i}$ diverges, one has to require
\be
\lambda = - \int_{-\infty}^{+\infty} \frac{dz'}{v_{\parallel}} \beta_{N}^{(k-1)}
 f_{i} \exp\left[ \int_{-\infty}^{z'} \frac{dz''}{v_{\parallel}} \beta_{i}
\right].
\ee
For $v_{\parallel}>0$, the integration constant can be determined by recalling 
that at upstream infinity there cannot be particles that have already suffered
$k$ scatterings and that move with $v_{\parallel}>0$, which implies $\lambda=0$.

It follows that the general solution of Eq.~\ref{eq:vla} for the partial functions
$f_{N}^{(k)}$ is:
\be
f_N^{(k)}= - \int_{z}^{+\infty} \frac{dz'}{v_{\parallel}} \beta_{N}^{(k-1)}
f_{i} \exp \left[ \int_{z}^{z'} \frac{dz''}{v_{\parallel}} \beta_{i} \right],
~~~v_{\parallel}<0,
\label{eq:fnk<}
\ee
and
\be
f_N^{(k)}= \int_{-\infty}^{z} \frac{dz'}{v_{\parallel}} \beta_{N}^{(k-1)}
f_{i} \exp \left[ \int_{z}^{z'} \frac{dz''}{v_{\parallel}} \beta_{i} \right],
~~~v_{\parallel}>0.
\label{eq:fnk>}
\ee

From Eq.~\ref{eq:vla}, and for $v_{\parallel}=0$, one easily obtains
\be
f_{N}^{(k)}(z,v_{\parallel}=0,
v_{\perp})=\left[\beta_{N}^{(k-1)}f_{i}/\beta_{i}\right]_{z,v_{\parallel}=0,v_{
\perp}}.
\ee 

The global solution of the Vlasov equation can now be written as the sum of all 
the partial functions:
\be
f_{N}(z,v_{\parallel},v_{\perp}) =
  \sum_{k=0}^{\infty} f_{N}^{(k)}(z,v_{\parallel},v_{\perp}).
\ee
In general, however, a good approximation to the solution is obtained when a
sufficient number of partial functions is taken into account. The needed  number
of partial functions is determined by the physical scales of the problem, as we
discuss below.

Let us now describe how all the relevant $f_N^{(k)}$ are computed and what kind
of approximations enter the calculation.

First, let us consider the ions. We assume that newly ionized particles reach
local thermal equilibrium with other ions in a time very short comparable with
all other time scales. This seems a reasonable assumption in a scenario in which
electromagnetic interactions lead to thermal equilibration. The ion distribution
is assumed to be a Maxwellian
\be
f_{i} (z,v_{\parallel},v_{\perp})  = \frac{n_{i}(z)}{\left[\pi \,
v_{th,i}(z)^{2} \right]^{3/2}} \exp\left[ -\frac{(v_{\parallel}-v_{i}(z))^{2} +
v_{\perp}^{2}}{v_{th,i}(z)^{2}}   \right],
\label{eq:max}
\ee
where $n_{i}(z)$, $v_{th,i}(z)$ and $v_{i}(z)$ are the density, thermal
velocity and bulk velocity respectively of ions at the position $z$. This form
of the ion distribution function allows us to calculate $\beta_{i}$ analytically
as described by \cite{Pauls95}:
\be
\beta_{i}(z,v_{\parallel},v_{\perp}) = m_{p} n_{i} (z) \sigma_{t} (U_{*}) U_{*},
\ee
where $U_{*}=\sqrt{\frac{4}{\pi}v_{th,i}^{2} + (v_{i}-v_{\parallel})^{2} +
v_{\perp}^{2}}$ and $\sigma_{t}=\sigma_{ce}+\sigma_{i}$ is the total cross
section for charge exchange and ionization. We checked that this approximation
leads to the correct result with an accuracy of order few percent. 

The assumption of rapid (local) thermalization of newly produced ions with
the bulk of ions deserves some discussion: this assumption is based on the fact
that the thermalization process proceeds through the excitation of
electromagnetic instabilities which are usually rather fast compared to the other 
processes involved. A {\it caveat} is due, however.
The fastest process is most likely that of isotropization of the
velocities of the newly created ions, which would make their effective thermal velocity comparable
with their initial bulk velocity. While downstream of the shock such effective
temperature is not too far from the temperature of thermal ions, in the upstream
plasma the newly born ions would end up having a temperature much higher than
that of the cold ions in the ISM. Whether the two populations can
indeed reach some sort of thermal equilibrium upstream within a convection time
to the shock is an open issue, and a rather difficult one. We are not aware
of any previous discussion of this problem in the existing literature \footnote{The possibility of non-Maxwellian proton distributions downstream of the shock as due to charge exchange reactions was presented by \cite{nonmax}.}, therefore we
adopt here the simplest assumption also made by all previous works on this
topic. It is however worth keeping in mind that the heating in the upstream
fluid predicted by our calculations may be somewhat overestimated
in case of partial rather than total local thermalization of ions.

As for the calculation of the coefficients $\beta_{N}^{(k)}$, this is the most
challenging part of the work from the point of view of computation time: these
are multi-dimensional integrals to be calculated on a multi-dimensional grid of
values of $(z,v_{\parallel},v_{\perp})$ and in general the functions
$f_{N}^{(k)}$ are far from being Maxwellian distributions, as discussed in
\S~\ref{sec:results}. We compute the $\beta_{N}^{(k)}$ following an approximate
but physically motivated procedure, which is discussed in detail in Appendix
\ref{app:appendix}.

We then need to describe the evolution of the ion component under the action of
charge exchange and ionization. As stressed above, ions are assumed to behave as
a fluid, therefore their dynamics is described by a set of conservation
equations which read:
\be
\frac{\partial}{\partial z} \left[ \rho_{i} v_{i} + F_{mass} \right]=0,
\label{eq:rh1}
\ee
\be
\frac{\partial}{\partial z} \left[ \rho_{i} v_{i}^{2} + P_{g,i}+ F_{mom} \right]=0,
\label{eq:rh2}
\ee
\be
\frac{\partial}{\partial z} \left[ \frac{1}{2}\rho_{i} v_{i}^{3} + 
\frac{\gamma_{g}}{\gamma_{g}-1}P_{g,i} v_{i}+ F_{en} \right]=0,
\label{eq:rh3}
\ee
where $F_{mass} = m_{p} \int d^{3} v v_{\parallel} f_{N}$, $F_{mom} = m_{p}
\int d^{3} v v_{\parallel}^{2} f_{N}$ and $F_{en} = m_{p}/2 \int d^{3} v
v_{\parallel} (v_{\parallel}^{2}+v_{\perp}^{2}) f_{N}$ are the fluxes of mass,
momentum and energy of neutrals along the $z$ direction. 

In these conservation equations we did not include radiative effects and recombination 
of neutrals which would profoundly change the dynamics of the shock region. This restricts
the range of applicability of this calculation to shocks moving with velocity $\gtrsim 500$
km/s, or more in general, to non-radiative shocks.

Notice that the terms on the right hand sides of Eqs.~\ref{eq:rh1}, \ref{eq:rh2} and \ref{eq:rh3} are
zero because mass, momentum and energy of the whole system (ions plus neutrals)
are conserved. It is also worth stressing that nowhere here we assume that
neutrals behave as a fluid. This point is in fact crucial in order to obtain the
correct solution of the problem: two and three fluid approaches fail to describe
the Physics correctly on scales smaller than the charge exchange length, as can
be easily understood given that the neutrals feel the presence of the shock only
indirectly, through their interactions with ions. In particular, it can be shown
that fluid approaches in a stationary situation leads in general to the
formation of a shock in the downstream neutral fluid: this is not what
happens in nature as confirmed by our calculations.

In practical terms our calculation is carried out by following an iterative
procedure: we start with neutrals and ions in thermal equilibrium at "upstream
infinity" (which translates, computationally, into {\it many interaction lengths
upstream of the shock}). The relative density of the two components is
parametrized through the ionization fraction. We start by fixing the profile of
density, pressure and velocity of the ions, so that an ordinary shock front is
formed at $z=0$ as predicted by solving the Rankine-Hugoniot equations for ions
alone. The corresponding $f_{N}^{(k)}$ are calculated by solving
Eqs.~(\ref{eq:fnk<}) and (\ref{eq:fnk>}) above, so that a mass, momentum and
energy flux of neutrals are determined. At this point, using the conservation
equations, Eqs.~(\ref{eq:rh1}), (\ref{eq:rh2}) and (\ref{eq:rh3}), we derive an
updated profile of the quantities describing the ions' dynamics, and the cycle
is restarted. 

Before embarking in the detailed explanation of the results of the
calculations, let us illustrate the basic Physics that is expected, starting
from the downstream region, where it is simplest. We will then proceed to
introduce the novel phenomenon of {\it return flux} which is of the highest
importance for astrophysical applications. 

The cold, fast neutrals from the upstream region penetrate the downstream region
where ions have been slowed down by the shock and correspondingly heated up. In
this situation the velocity difference between ions and neutrals ignites the
processes of charge exchange and ionization (they also occur upstream, see
below). A fast neutral that becomes an ion through charge exchange
and/or ionization delivers momentum and energy to the ions. In the absence of
ionization, one would expect that a large number of charge exchange events leads
to thermalization between ions and neutrals. 
On scales much larger than the interaction lengths for the relevant processes we 
can think of using a "black box" approximation, namely of writing the mass, momentum and energy 
conservation equations as between upstream and downstream infinity (subscript $1$ and $2$ respectively). 
We would then find that the final state (at downstream infinity) corresponds to a total compression 
factor $R_{tot}$ that is the one appropriate for a system with total density equal to the sum of the densities 
of charged and neutral atoms ($\rho_i$ and $\rho_N$ respectively). In the presence of ionization only charged 
particles will be left at downstream infinity, with a density $\rho_{i,2}=
R_{tot}\left(\rho_{i,1}+\rho_{N,1}\right)$.

As we already mentioned, the heating of neutrals through charge exchange is a
phenomenon of crucial importance in that it affects the width of Balmer lines 
emitted by these hot neutrals. The line width is, in turn, a direct measurement
of the temperature of ions downstream of the shock, at least on a spatial scale
which is comparable with the excitation one (which is also comparable with the
ionization scale). 

Neutrals that suffer one or few charge exchanges downstream have a very
anisotropic distribution function. Charge exchange with ions in the part of the
distribution function with $v_{\parallel}<0$ generates new neutrals which move
against the stream of incoming particles in the shock frame. 

A few considerations are useful at this point: for a strong shock, if not modified by CRs, the thermal
distribution of ions is centered around a bulk velocity which is $\sim V_{sh}/4$
and has a spread of the order of $\sim (3/8)^{1/2} V_{sh}$. It is therefore easy
to visualize that in the shock frame there are many ions that have a
negative speed in the $z$ direction, namely that move against the stream. As
long as these particles are charged this does not represent a problem, in that
their gyration radius is very small and they remain behind the shock \footnote{In fact, even a small fraction of ions might recross the shock, as discussed by \cite{eichler}, and eventually trigger the beginning of the injection process.}. However,
following a charge exchange event, an ion with $v_{\parallel}<0$ becomes a
neutral that keeps moving in the same direction. 

These neutrals have a finite probability of reaching upstream because
again they are insensitive to the electromagnetic fields at the shock. When one such
neutral happens to cross the shock and undergo a new interaction in the
upstream, the associated deposition of energy and momentum will cause the
heating of the upstream fluid.
This is what we will refer to as the \emph{neutral return flux}, a phenomenon
that could not be found in previous calculations, either because carried out in
the fluid approximation or because concentrated on the downstream part of the
plasma. 
Energy and momentum deposition upstream occur on a spatial scale which is the
minimum between the charge exchange and the ionization scales. Whether one or
the other dominates depends mainly upon the shock velocity. The relative
velocity between the returning neutral and an ion is of order $\sim 2 V_{sh}$. 
For shocks with $V_{sh}\ga 3000$~km~s$^{-1}$ the relative velocity is such that
the cross section for charge exchange is suppressed and the returning neutrals
heat up the upstream fluid on the ionization scale (see Fig.~\ref{fig:cross}). 
For slower shocks the energy deposition takes place on the scale of charge
exchange, since ionization is a threshold process in terms of relative speed.
However one should also keep in mind that for high velocity shocks for most
neutrals the first interaction downstream is of ionization type, therefore in
this case the return flux is suppressed.

We stress that, since the upstream gas is cold ($T\sim 10^{4}$ K), even a
relatively small energy deposition may lead to a large increase in the ion
temperature and a corresponding reduction in the Mach number of the plasma
immediately upstream of the shock, which is thereby weakened. In \S~\ref{sec:cr}
we will describe how the return flux may affect the spectra of CRs
accelerated at such type of shocks.

\section{Results}
\label{sec:results}

In this section we illustrate the main results of our kinetic calculations for a
benchmark case with shock speed $v_{s}=2000$~km~s$^{-1}$, total density of
$0.1$~cm$^{-3}$ and 50 \% ionization fraction (we specify the values of these
parameters whenever they differ from the above). The temperature of the gas at
downstream infinity is always assumed to be $T=10^{4}$~K in order to be
compatible with the presence of neutrals. The temperature of neutrals is assumed
to be equal to that of ions, so at upstream infinity the two components are in
thermal equilibrium. In this range of parameters, typical of SNRs that are expected to play
an important role for particle acceleration, the shock is non-radiative, therefore the calculations
presented in the section above fully apply.

\subsection{Dynamical properties of neutrals and ions}
\label{sec:hydro}
\begin{figure}
\begin{center}
{
\includegraphics[width=0.47\textwidth]{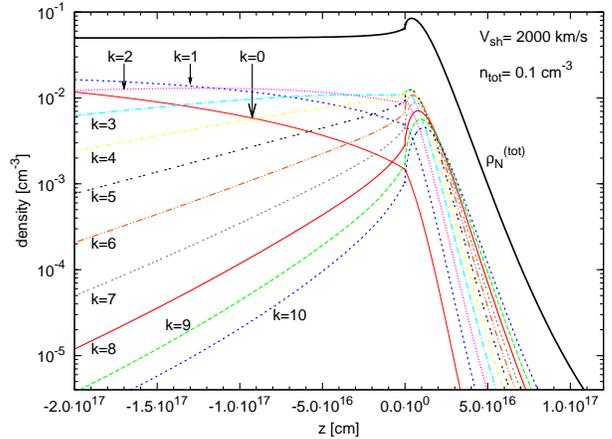}}
\end{center}
\caption{Density of particles in the individual distributions $f_{N}^{(k)}$with
$k=0,~1,~2,...$ as indicated. The thick solid line is the density of neutrals as
a function of the distance from the shock upstream ($z<0$) and downstream
($z>0$).}
\label{fig:nk}
\end{figure}

In Fig.~\ref{fig:nk} we show the density of particles in each $f_{N}^{(k)}$,
namely $\int d^{3}v f_{N}^{(k)}(\vec v)$, as a function of the distance from the
shock, both upstream ($z<0$) and downstream ($z>0$). 
As expected, the density of particles that did not suffer any charge exchange
($k=0$) decreases monotonically with distance from upstream infinity. Clearly
the concept of upstream infinity does not have physical relevance for us: moving
it further away from the shock simply leads to requiring a larger number of
$f_{N}^{(k)}$ to describe the total distribution of neutrals. 
Moving it closer to the shock surface is also possible, provided the neutrals
that return from downstream after one charge exchange event become ionized or
suffer another charge exchange within the distance associated with the {\it
downstream infinity} boundary. The density associated with the $f_{N}^{(k)}$
with $k>0$ is not a monotonic function of the position because each distribution
not only receives a contribution from the neutrals that have suffered $(k-1)$
charge exchange reactions, but at the same time is also deprived of particles
because of additional charge exchange and ionization interactions. One can see
this phenomenon in Fig.~\ref{fig:nk} by comparing the contribution of $k=0$ and
$k=1$: the density of particles with $k=1$ increases with $z$ within one
interaction length of charge exchange, and then starts decreasing as a
consequence of additional charge exchanges, which correspondingly contribute to
the $f_{N}^{(k)}$ with $k>1$. Once the neutrals penetrate the downstream region
they feel a larger density of ions and a larger velocity difference. 
The total density of neutrals ($\rho_N$) is plotted in Fig.~\ref{fig:nk} as a
solid thick line. Approaching the shock from upstream, $\rho_N$ increases
slightly because of the contribution of the return flux, namely of neutrals that
come from the downstream region after a hot ion experienced a charge exchange
thereby becoming a fast neutral that may move against the flow in the upstream.
Moving from the shock vicinity to far upstream, the total density of neutrals
becomes constant, because the returning neutrals disappear progressively,
either due to ionization or to charge exchange.
In the far downstream region the density of neutrals decreases again because of
ionization.  

\begin{figure}
\begin{center}
\includegraphics[width=0.47\textwidth]{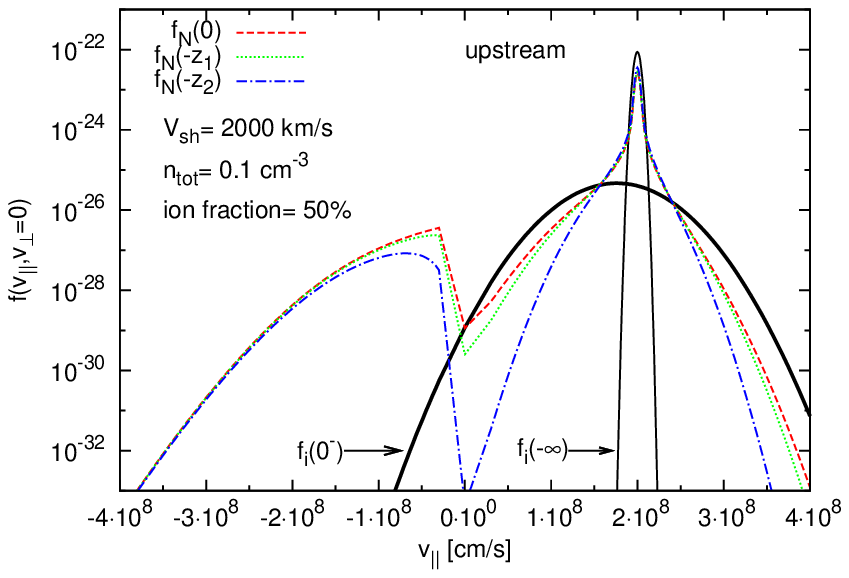}
\includegraphics[width=0.47\textwidth]{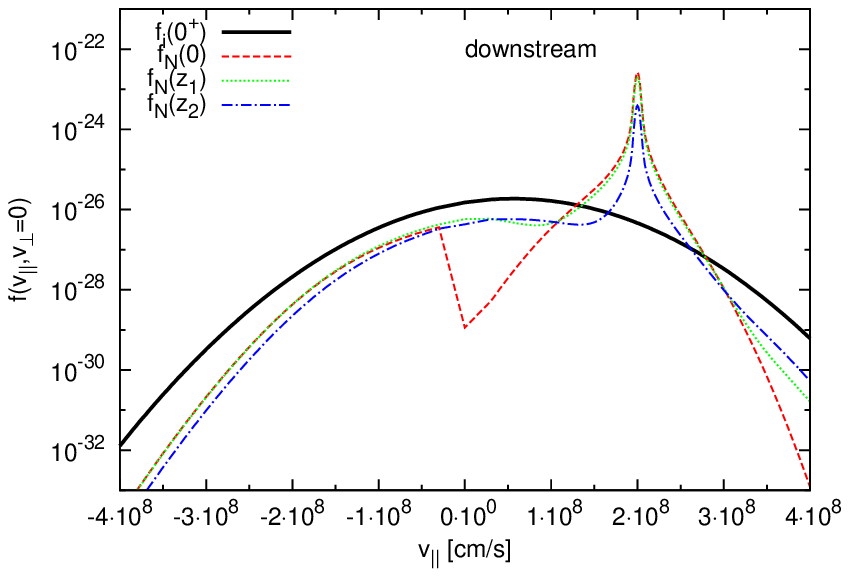}
\end{center}
\caption{{\it Left Panel:} Distribution function of neutrals at the shock
($z=0$, dashed) and at the locations $z_{1}=2.1\times 10^{15}~\rm cm$ (dotted
line) and $z_{2}=1.3 \times 10^{16}~\rm cm$ (dash-dotted line). The thick solid
line is the distribution of ions immediately upstream of the shock while the thin
solid line is the ion distribution at upstream infinity.
{\it Right Panel:} as in the left panel but for the downstream region. The thick
solid line is the distribution of ions immediately downstream of the shock.}
\label{fig:fn}
\end{figure}

The distribution functions in phase space that result from our calculations are
by no means Maxwellian functions, and contrary to previous calculations (e.g.
see \cite{HengA07}) they are computed at any point in space rather than being
volume averaged. 
It is not easy to summarize in a plot the global properties of the functions
$f_{N}$ since they depend on the position $z$ and on velocity $\vec v =
(v_{\parallel},\vec v_{\perp})$.
In Fig.~\ref{fig:fn} we show $f_{N} (z,v_{\parallel},v_{\perp}=0)$ at two
locations $z_{1}=\pm 2.1\times 10^{15}~\rm cm$ and $z_{2}=\pm 1.3 \times
10^{16}~\rm cm$ (the minus and plus signs refer to upstream and downstream
respectively). The left and right panels show our results for the upstream
and downstream section respectively.

The thin and thick solid lines (both Maxwellians) in the left panel are the ion
distribution at upstream infinity and immediately upstream of the shock
respectively. The long-dashed line is the distribution of neutrals at the shock
location (since neutrals do not feel the shock directly their distribution is
the same across the shock front). The short-dashed and dash-dotted lines refer
to the distribution of neutrals at locations $-z_{1}$ and $-z_{2}$. At large
distances from the shock the distribution of neutrals is identical to that of
ions with temperature $10^{4}$ K. In previous calculations of the structure of
shocks in partially ionized media this situation was assumed to extend to the
shock itself, namely nothing was happening in the upstream plasma. We find here that
this is not a good description of reality because of the neutrals return flux.
This component is clearly visible in the left panel of
Fig.~\ref{fig:fn}, in the region $v_{\parallel}<0$. The
most important effect induced by this return flux of neutrals is the heating of
upstream ions. Due to charge exchange interactions between these hot upstream
ions and cold upstream neutrals (the ones with $v_\parallel>0$), the
distribution function of the latter gets broadened, compared to the narrow
Maxwellian at upstream infinity (see short-dashed and dash-dotted lines in the
left panel of Fig.~\ref{fig:fn}). Clearly far enough upstream the return flux
($v_{\parallel}<0$) disappears as a result of ionization and additional charge
exchange reactions, so that, as we stressed above, the distribution of neutrals
at upstream infinity sits on the ion distribution at $T=10^{4}$ K.

The situation downstream of the shock is somewhat simpler and is illustrated in
the right panel of Fig.~\ref{fig:fn}. The distribution of neutrals at the shock
(dashed line) is the same as in the left panel, but the neutrals at
$z_{1}$ and $z_{2}$ developed a very broad distribution, roughly with the same
width of the ions' distribution immediately behind the shock (thick solid line).
This is the result of efficient charge exchange between the cold ($T=10^{4}$ K)
neutrals and the hot ions behind the shock. Nonetheless, one can also see that
at the distances $z_{1}$ and $z_{2}$ considered here there is still a leftover
narrow distribution of particles from upstream. Their contribution to the total
number and energy is however rather small in the downstream plasma. Moving
towards downstream infinity the neutrals disappear as a result of ionization,
although for slow moving plasmas this phenomenon can occur very far from the
shock because of the small ionization cross section. 

\begin{figure}
\begin{center}
\includegraphics[width=0.47\textwidth]{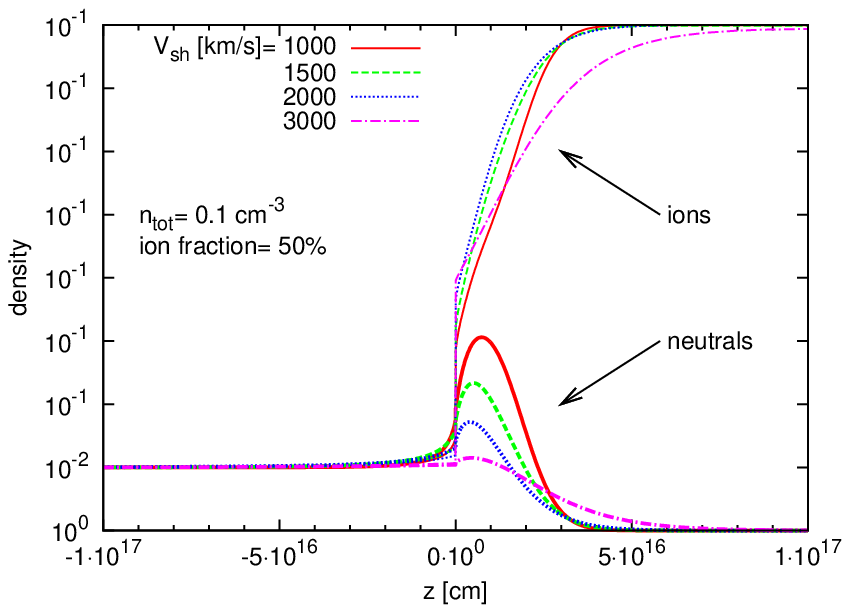}
\includegraphics[width=0.47\textwidth]{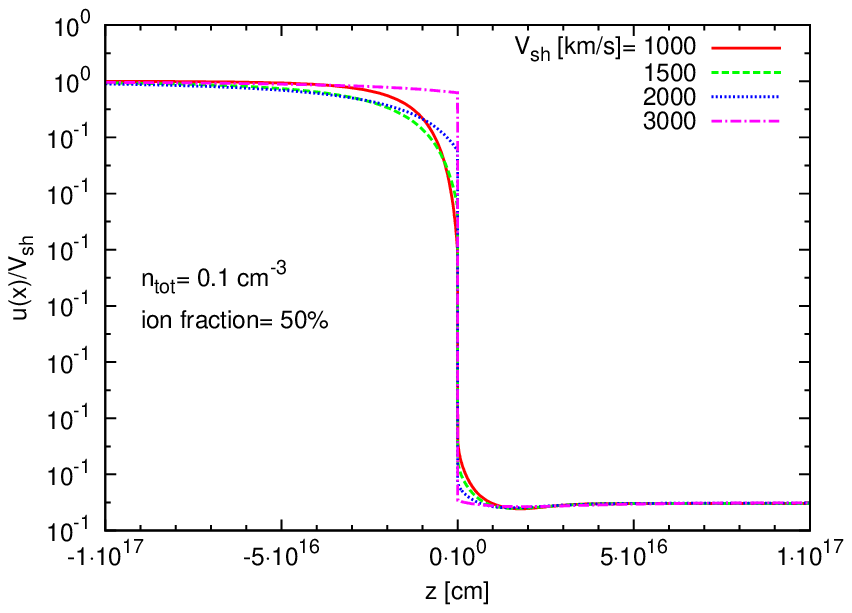}
\end{center}
\caption{{\it Left Panel:} Density of neutrals (lower curves) and ions (upper
curves) as a function of the distance from the shock upstream and downstream,
for different values of the shock velocity. 
{\it Right Panel:} Normalized ion velocity for different values of the shock
speed.}
\label{fig:hydro}
\end{figure}

The role of the return flux is better illustrated by the dynamical quantities
associated with ions. In Fig.~\ref{fig:hydro} we show the density (left) and
velocity (right) of the ion plasma as a function of the position across the
shock ($z=0$). The different curves refer to different values of the shock
velocity as labelled. One can see that the density stays constant upstream until
very close to the shock front, where the return flux becomes important. In this
region, the fast neutrals from downstream deposit energy into the ion plasma
through both ionization and charge exchange, thereby heating the gas. These
reactions also deposit momentum in the $-z$ direction, thereby slowing
down the ion plasma (right panel). The combination of this induced deceleration and 
of the ionization process causes the ion density to increase (left panel). One final comment 
is deserved by the right panel of Fig.~\ref{fig:hydro}: here we see that with decreasing shock 
velocity between 3000 and 1500~km~s$^{-1}$ the precursor becomes progressively more pronounced. 
This trend suddenly changes for $V_{sh}$=1000~km~s$^{-1}$, when the precursor becomes shorter
and steeper: this is due to the fact that in this case the return flux is destroyed by charge-exchange 
events rather than by ionization, and those occur on a shorter scale.

\begin{figure}
\begin{center}
{
\includegraphics[width=0.47\textwidth]{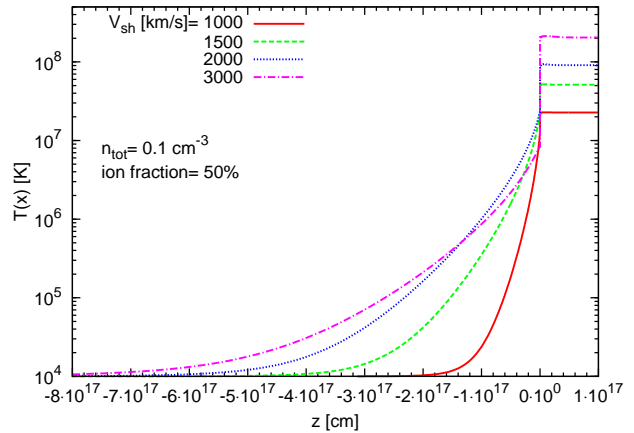}
}
\end{center}
\caption{Temperature of ions as a function of position around the shock, for
different values of the shock speed.}
\label{fig:temp}
\end{figure}

It is important to realize that the whole system of neutrals and ions on very
large scales, namely between upstream infinity and downstream infinity, must
behave as a black box in which the standard conservation equations must apply
(a generalized version of this statement can also be written in the presence of CRs).
Since for $T=10^{4}$~K we have a sonic Mach number $M = 85~V_{sh}/(1000\,{\rm
km~s}^{-1})\gg1$, the shocks we are dealing with are indeed strong: therefore, the velocity
of ions and neutrals, at upstream infinity, has to be $\sim 4$
times larger than the velocity of the far downstream plasma, made of ions only
because of ionization. This fact is clearly visible in the right panel of
Fig.~\ref{fig:hydro}. From mass conservation one immediately obtains that 
\be
\rho_{i} V_{sh} + \rho_{N} V_{sh} = \frac{V_{sh}}{4} \rho_{F} \to \rho_{F} = 4 \left( 
\rho_{i}  + \rho_{N} \right) ,
\ee
where $\rho_{F}$ is the density in the far downstream plasma.
For equal densities of neutrals and ions at upstream infinity,
$\rho_{F}/\rho_{i}=8$. This is clearly visible in the left panel of Fig.~\ref{fig:hydro}, 
where the upstream density is $0.05~\rm cm^{-3}$ and the
density at downstream infinity $\rho_{F}$ is $0.4~\rm cm^{-3}$, independently of the shock
velocity. Increasing the shock velocity leads to reducing the relative
importance of the return flux, as illustrated in Fig.~\ref{fig:temp}, where we
plot the temperature of ions as a function of location around the shock. 

All curves start at $T=10^{4}$ K at upstream infinity, but it is clear that, on
the way to the shock, ions are heated to much higher temperatures due to the
effect of the return flux described above. The fact that the gas temperature is
of order $\sim 10^{7}$ K immediately upstream of the shock leads to a dramatic
reduction in the ion Mach number and therefore in the compression factor at the
shock which is appreciably smaller than the canonical value of $4$, especially
for low shock velocities (below $\sim 3000$~km~s$^{-1}$). 

In the perspective of discussing the implications of these physical processes
for diffusive particle acceleration at shocks, it is worth stressing that the
effect of neutrals on the shock structure is that of producing a precursor
upstream of the shock, but that the nature of this precursor is totally
unrelated to the CR induced precursor that is found in non-linear
theories of particle acceleration  \cite[see e.g][for a review]{reviewNLA}.
In analogy with CRs, the neutrals can carry information from downstream to
upstream of the shock. They then deposit such information on spatial scales which are
fully determined by the cross sections of charge exchange and ionization. In the
case of CR-induced precursors, the plasma upstream is slowed down by the pressure of accelerated particles, and the spatial scale of the precursor is determined by the diffusion properties of the medium and by the spectrum of accelerated particles (in general the precursor is larger whenever the spectra are harder, a sign of more efficient acceleration). The spatial extent of the neutral-induced precursor is basically fixed by the cross sections for charge exchange and ionization. The heating of the plasma in a CR-induced precursor is due to adiabatic heating and to turbulent heating \citep{be99}. 
The latter depends on unknown details of wave damping on the gas, and its efficiency can only be parametrized. The process cannot be too effective otherwise the wave amplification that is responsible for effective diffusion upstream is inhibited. In a neutral-induced precursor, the heating is due to energy and momentum deposition of ions produced in charge exchange and ionizations reactions of returning neutrals upstream. The only uncertainty here is due to the unknown rapidity of ion assimilation in the thermal plasma (see also the discussion below).

\subsection{Acceleration of test particles in partially ionized media}
\label{sec:cr}

In the context of diffusive shock acceleration the spectrum of test particles
accelerated at the shock is a power law $N(E)\propto E^{-\gamma}$ with a slope
$\gamma$ fully determined by the compression factor at the shock:
\be\label{eq:gamma}
\gamma=\frac{r+2}{r-1}.
\ee
For a strong shock (sonic Mach number $M \gg 1$), if not modified by CRs, 
the compression factor $r\to 4$
and the spectrum reaches its asymptotic shape $N(E)\sim E^{-2}$. 

\begin{figure}
\begin{center}
\includegraphics[width=0.47\textwidth]{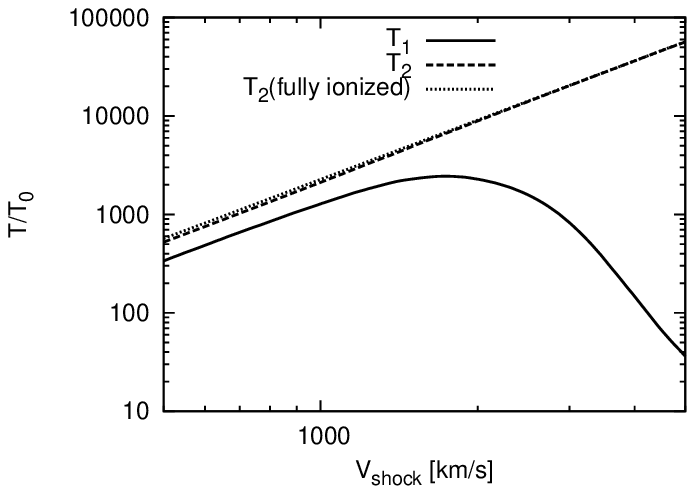}
\includegraphics[width=0.47\textwidth]{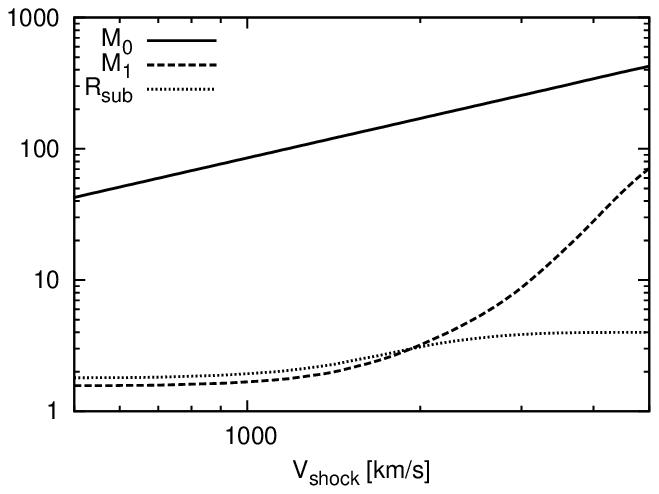}
\end{center}
\caption{{\it Left Panel:} Temperature of ions immediately upstream ($T_{1}$)
and downstream ($T_{2}$) of the shock. The dotted line is the downstream
temperature in the absence of neutrals. {\it Right Panel:} Mach number of the
ions fluid at upstream infinity (solid line) and immediately upstream of the
shock (dashed line). The dash-dotted line shows the compression factor at the
subshock in the presence of neutrals.}
\label{fig:mach}
\end{figure}

As discussed in \S \ref{sec:hydro} the presence of neutrals induces the
formation of a precursor upstream of the shock front. The ion temperature
immediately upstream of the shock may become 2-3 order of magnitude larger than
the temperature at upstream infinity, hence the Mach number at the shock is much
reduced. The importance of this effect depends upon the shock velocity. In the
left panel of Fig.~\ref{fig:mach} we show the temperature immediately before
($T_{1}$) and behind ($T_{2}$) the shock as a function of the shock
velocity. The dotted line illustrates the downstream ion temperature in the
absence of neutrals.

The temperature immediately behind the shock is basically the same with or
without neutrals. This fact can be understood, in a qualitative way, by considering
a "black box" description of the whole system, namely writing the conservation 
equations for mass, momentum and energy between upstream and downstream infinity,
while ignoring the detailed physics in between. When this is done, the ion temperature at 
downstream infinity is fixed, independent of the presence of neutrals. If one further considers
that charge exchange processes in the downstream have very little effect on the temperature
of ions, the above result is readily interpreted.

What is most impressive is that the temperature of ions upstream grows to very large values, $T_{1}\sim 10^{6}-10^{7}$ K due to the presence of neutrals.
We stress that such a heating may be much stronger than the one generated by any other mechanism, 
like for instance the turbulent heating due to the damping of Alfv\'en waves, which may be expected to be effective in CR-modified shocks \cite[e.g.][and references therein]{be99,lungo}.
The reason is that the reservoir of energy to be damped by Alfv\'en heating is the energy in the form of magnetic turbulence (typically less than 1 per cent of the bulk energy upstream), while here the neutral return flux itself accounts for a potentially large fraction of the bulk pressure (see the right panel in fig.~\ref{fig:hydro}). 

$T_{1}$ grows with shock velocity until it reaches a maximum around $\sim 2000$~km~s$^{-1}$,
and decreases for faster shocks. This trend reflects the physical essence
of the return flux: for shock velocities smaller than $\sim (2-3) \times 10^{3}$~km~s$^{-1}$
neutrals entering the downstream plasma are most likely to suffer a charge
exchange interaction, and the resulting fast neutral has a finite probability of
having $v_{\parallel}<0$ thereby contributing to the return flux. As we already mentioned,
in this situation the ion plasma upstream gets heated and a precursor is formed. On the
other hand for $V_{sh}>(2-3) \times 10^{3}$~km~s$^{-1}$, ionization occurs before charge
exchange and the return flux is correspondingly suppressed. This explains the
decline of $T_{1}$ in the left panel of Fig.~\ref{fig:mach} in the high shock
speed region.

In the right panel of Fig.~\ref{fig:mach} we plot the Mach number at upstream
infinity ($M_{0}$, solid line), the Mach number of ions immediately upstream of
the shock ($M_{1}$, dotted line) and the compression factor at the (sub)shock,
$R_{sub}$ (dashed line), as functions of the shock velocity. For all values of
$M_{0}$ the compression factor that would be derived in the absence of neutrals
is $\sim 4$, but the action of neutrals is such that the compression factor
drops to values below $2$ for $V_{sh}<1500$~km~s$^{-1}$ and gradually grows to 4,
which is reached however only for $V_{sh}\ga 3000$~km~s$^{-1}$. 

Even on a qualitative basis it is clear that the presence of neutrals, by affecting the 
compression factor at the shock, will also affect the spectrum of test particles accelerated 
at shocks: more specifically, the latter will become steeper than standard test particle
theory would predict for a high Mach number shock. It is also clear that the
spectrum of accelerated particles must be concave to some extent because the
compression factor experienced by low energy particles is closer to $R_{sub}$, while
higher energy particles experience a compression factor closer to the standard one,
$\sim 4$.

In order to estimate this effect we introduce the energy-dependent compression factor 
\begin{equation}
 R(E) = u_{E}^{(1)} / u_{E}^{(2)},
\end{equation}
where 
\begin{equation}
 u_{E}^{1,2} = u_{1,2} + \frac{1}{N_{0}(E)} \int dx \frac{d u }{d x} N(E,x) 
\end{equation}
represents an effective fluid velocity upstream (1) and downstream (2) of the
shock, as experienced by particles with energy $E$, $N_{0}(E)$ is the spectrum of
particles at the shock location, and $u_{1}$ ($u_{2}$) is the fluid velocity
immediately upstream (downstream) of the shock. 
The spatial integral is extended to upstream infinity for $u_{E}^{(1)}$ and to
$D(E)/u_{2}$ downstream for $u_{E}^{(2)}$, where $D(E)$ is the diffusion
coefficient for a particle with energy $E$. The reason is that the downstream
region in principle extends to infinity, but the particles that can return to
the shock due to diffusion are only those that reside within a region of size
$D(E)/u_{2}$ downstream. All the particles in the upstream region are eventually
advected towards the shock front. Here the diffusion coefficient is assumed to
be Bohm-like with a magnetic field of $\sim 10~\mu$G upstream and $\sim
\sqrt{11} \times 10~\mu$G downstream (formally the factor $\sqrt{11}$ holds only
for compression of a turbulent magnetic field at strong shocks with compression
factor 4, but this is not very important in this context). 

\begin{figure}
\begin{center}
\includegraphics[width=0.47\textwidth]{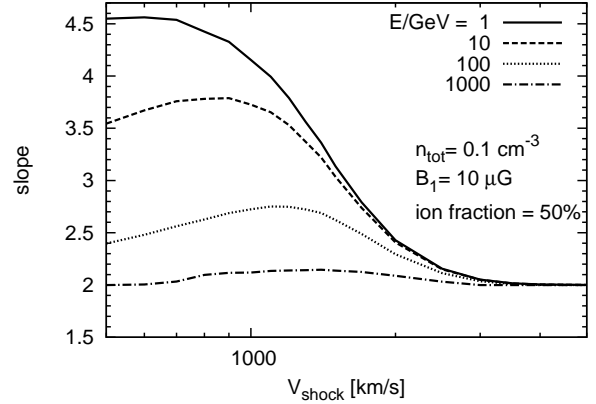}
\end{center}
\caption{Slope of the spectrum of accelerated test particles for
$E=1,~10,~100,~1000$ GeV, as a function of the shock speed.}
\label{fig:slope}
\end{figure}
   
The compression factor $R(E)$ provides an estimate of the actual compression
factor experienced by particles with energy $E$. The slope of the spectrum is
therefore defined as $\gamma(E)=\frac{R(E)+2}{R(E)-1}$ (see Eq.~\ref{eq:gamma}) 
and plotted in Fig.~\ref{fig:slope} for four values of the particle energy
($E=1,~100,~100,~1000$ GeV) as a function of the shock speed. 

Very high energy particles ($E=1$ TeV, dotted line) sample almost the entire
large-scale structure of the shock so that for them the effective compression
factor is close to 4. The corresponding spectral slope varies between $2$ (for
very slow and very fast shocks) and $2.1$ for $V_{sh}\sim 1500$~km~s$^{-1}$. 

For particles with $E=1$ GeV, the slope is considerably affected by the
presence of neutrals, becoming as large as $\sim 4.5$ for $V_{sh}\la
500$~km~s$^{-1}$. The slope approaches the canonical value of $2$ only for
$V_{sh} \ga 4000$~km~s$^{-1}$. The effect of neutrals is very evident also for
100 GeV particles (short-dashed line): the slope gets as steep as $\sim 2.7$ for
$V_{sh}\sim 1000$~km~s$^{-1}$, and is always larger than 2.3 for $V_{sh}\la
2500$~km~s$^{-1}$.

\begin{figure}
\begin{center}
\includegraphics[width=0.47\textwidth]{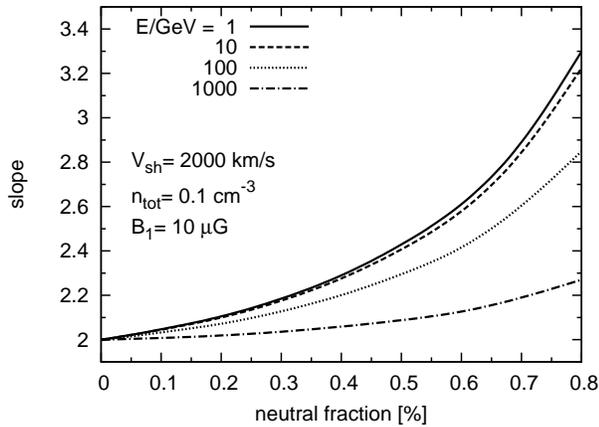}
\end{center}
\caption{Slope of the spectrum of accelerated test particles for 
$E=1,~10,~100,~1000$ GeV, as a function of the fraction of neutrals, for a shock
with velocity $V_{sh}=2000$~km~s$^{-1}$ and a total density at upstream infinity
$n=0.1~\rm cm^{-3}$.}
\label{fig:slopexi}
\end{figure}

These results clearly show how the spectrum of accelerated particles is affected
in a very important way by the presence of neutrals for ionization fraction
of $50\%$ (our benchmark case) and shock velocity $\la 4000$~km~s$^{-1}$.
In Fig.~\ref{fig:slopexi} we plot the spectral slope of test particles for
$E=1,~10,~100,~1000$~GeV, $V_{sh}=2000$~km~s$^{-1}$, $n=0.1$~cm$^{-3}$ as a
function of the fraction of neutrals.

A departure of the spectral slope of accelerated particles from the canonical 
value of $2$ is observed as soon as the neutral fraction is non-vanishing. 
The spectrum becomes especially steep at low energies, since these particles
probe spatial scales that are entirely contained within the precursor induced
by the return flux of neutrals rather than the global extent of the system.
For a neutral fraction $\sim 0.8$ even the spectral slope at $\sim 1$ TeV is
$\sim 2.3$. 

It is worth comparing the spectral steepening induced by the presence of
neutrals with that induced by non-linear effects in particle acceleration. Very
efficient acceleration does lead to steep spectra at energies below a $\sim 10$
GeV, as a consequence of the formation of a pronounced CR-induced precursor: the
steepening is caused by the fact that low energy particles only experience the
compression factor at the subshock, which is $<4$ if acceleration is efficient. 
The escape of particles at the highest achievable momenta makes the total
compression factor $>4$ (similar to a radiative shock), so that the particle
spectra at energies above $\sim 20$ GeV are harder than $E^{-2}$. While both the
test particle and the non linear theory of diffusive particle acceleration at
SNR shocks lead to predict spectra with high energy slope $\le 2$, observations
by Fermi and AGILE in the GeV band, and by HESS, VERITAS and MAGIC in the TeV
band, showed compelling evidence for gamma-ray spectra typically in the range
$E^{-2.2}-E^{-2.4}$ for shell-like SNRs (with the exception of RX J1713.7-3946
and Vela Jr.) and even steeper ($E^{-2.7}-E^{-2.9}$) for SNRs interacting with
(partially neutral) molecular clouds \citep[see e.g.][for a review and a wider
discussion]{dam11}. 

It has been pointed out, for instance by \cite{ziraptu} and \cite{lungo}, that
spectra steeper than $E^{-2}$ (and correspondingly lower efficiencies of
particle acceleration) can be obtained even in the context of the non-linear
theory of diffusive shock acceleration if the velocity of the scattering centers
is taken into account. However it is worth recalling that in these cases the
results are strongly dependent upon the detailed nature of the waves and on
their helicity \cite[see also][]{centers}: in principle the same effect may lead
to harder spectra rather than to a steepening. On the other hand, the neutral
return flux induces a precursor whose length scale (the
charge-exchange/ionization mean free path) is typically much larger than the
diffusion length of few GeV particles, thereby potentially affecting several
decades of the CR spectrum up to multi-TeV energies, as illustrated in
Fig.~\ref{fig:slopexi}. The SNRs from which we detect gamma rays of possible
hadronic origin are expected to accelerate particles with efficiencies of order
$\sim 10\%$, for which the non linear effects discussed above cannot be
neglected. In the absence of a theory that takes into account both the CR
modification and the neutral return flux, one cannot claim that the problem of
steep spectra is solved by the presence of neutrals, but it is clear that the
role of neutral atoms inside the accelerator may be very important in making the
present discrepancy between theory and observations milder.
\newline

\section{Conclusions}
\label{sec:conclude}

The structure of a collisionless shock wave is profoundly affected by the
presence of neutral atoms in the medium in which the shock propagates. The
coupling between the shocked ions and the neutrals occurs through the processes
of charge exchange and ionization and leads to strong modifications of the shock
structure.

Our calculations are based on a novel procedure that allows us to solve
semi-analytically the Vlasov equation describing the behaviour of neutrals and
the fluid equations describing the ionized plasma. Both ions and neutrals are
evolved from upstream infinity to downstream infinity and the structure of the
shock is calculated iteratively. The main physical phenomenon that we
discovered, and the one that is responsible for all the interesting effects
discussed here, is the existence of a return flux of neutrals in the upstream: a
cold neutral crossing the shock from upstream may suffer a charge exchange
reaction with a hot ion in the downstream; some of these interactions will
involve ions that are moving towards the shock (negative value of the velocity
component parallel to the shock normal, $v_\parallel$); when this happens, the
ion is transformed into a neutral that keeps moving with the same velocity,
$v_{\parallel}<0$, thereby recrossing the shock and reaching upstream. The
return flux of neutrals created in this way is then dissipated upstream through
additional charge exchange and ionization reactions. This leads to heating of
the upstream ions and to the consequent decrease of the shock Mach number with
respect to the value at upstream infinity. We find that in some cases the Mach
number drops to values of order $\sim 2$ immediately before the shock. This
implies that: 1) a precursor is induced upstream of the shock by the neutral
return flux; 2) the compression factor at the shock is lowered much below the
standard value of $4$ that applies to strong shocks, if not modified by CRs. 

There is an intrinsic velocity scale in the problem, which is of order
$2000-3000$~km~s$^{-1}$ established by the cross sections of ionization and
charge exchange: for shocks with $V_{sh}<(2-3)\times 10^{3}$~km~s$^{-1}$, a
neutral that crosses the shock is more likely to suffer a charge exchange
reaction with an ion downstream rather than being ionized. In these conditions a
flux of neutrals with $v_{\parallel}<0$ is created and the shock is profoundly
modified. At higher shock velocities the neutral gets ionized before it suffers
charge exchange, therefore the return flux is suppressed and the shock structure
is not affected appreciably. The spatial scale of the precursor induced by the
return flux is determined by the shortest between the ionization and the charge
exchange interaction lengths. This scale is again a function of shock velocity:
the relative velocity between an incoming ion and a returning neutral is of
order $\sim 2 V_{sh}$. If $2 V_{sh}<(2-3)\times 10^{3}$~km~s$^{-1}$ the main
process for dissipation of the return flux upstream is charge exchange,
otherwise energy and momentum are deposited through ionization. 

In our calculations neutrals are described through the Vlasov equation,
therefore we can determine their distribution function at any location upstream
and downstream. These distributions are not Maxwellian in shape: upstream of the
shock the particle distribution function is bimodal, with a roughly Maxwellian
peak that describes the neutrals that did not suffer charge exchange reactions
yet, and a broader component in the region $v_{\parallel}<0$ that disappears
while moving towards upstream infinity. In the downstream region the
distribution function competing to neutrals that have not suffered any charge
exchange rapidly vanishes and is replaced by a broad component, made of hot
neutrals and reflecting the charge exchange reactions with hot ions. Moving
towards downstream infinity all neutrals eventually disappear because of
ionization. The temperature of ions upstream is a strong function of distance
from the shock, as a consequence of the return flux. Downstream the ion
temperature varies little. 

The implications of the formation of a neutral-induced precursor  for particle
acceleration at collisionless shocks are of considerable interest. The spectrum
of test particles resulting from diffusive shock acceleration is determined by
the compression factor the particles experience while they diffuse in the region
surrounding the shock. We find that in the presence of neutrals this compression
ratio, even in the case of a high Mach number shock, is smaller than the
canonical value of $4$. It follows that for all particles for which the
diffusion length upstream, $\sim D(E)/V_{sh}$ is shorter than the precursor
length, the spectrum is steeper than $E^{-2}$. Our calculations provide a
quantitative confirmation of this qualitative expectation: assuming conditions
typical of supernova remnant shocks, we obtain that the spectrum in the $1-10$
GeV range may become as steep as $\sim E^{-4}$ for $V_{sh}=1000$~km~s$^{-1}$; at
100 GeV the spectrum is still $\sim E^{-2.5}$ and flattens to $\sim E^{-2.2}$
for TeV energies. The precursor weakens at larger shock velocities and the
particle spectrum becomes progressively less deviant: for $V_{sh}=2000$~km~s$^{-1}$,
the spectrum is always between $E^{-2.4}$ and $E^{-2.1}$. At even
larger shock velocities the standard results are reproduced. 

These results are obtained by assuming that the accelerated particles behave as
test particles. Their dynamical reaction is expected to become important in realistic
situations and this leads to a bunch of new effects that will be discussed in
a forthcoming paper. Just one instance of the complications that arise is that
the CR induced precursor slows down the ions with
respect to neutrals, thereby establishing a velocity difference that triggers
charge exchange far upstream of the shock. This effect greatly modifies the
shock structure as we discuss in a forthcoming paper, where the
calculations illustrated here will be generalized to the case of non-linear
shock acceleration. There, we will also illustrate the results of our
calculations on the width of Balmer lines in shocks where efficient CR
acceleration takes place. 

\acknowledgments
We are grateful to Bill Matthaeus for numerous useful discussions on the
physics of the solar wind. We are also grateful to the referee for his/her comments that 
helped improve this paper. This work was partially funded through grant
ASI-INAF I/088/06/0 and PRIN INAF 2010. The research work of  D.C. was partially
supported by NSF grant AST-0807381.

\appendix
\section{Calculation of $\beta_{N}^{(k)}$}
\label{app:appendix}

The calculation of the scattering rates $\beta_{N}^{(k)}$ has to be carried out
in an approximate way since the calculation of the integrals in $d^{3} \vec v$
on the entire three dimensional grid $(z,v_{\parallel},v_{\perp})$ leads to
exceedingly long computation times. The expressions for $f_{N}^{(k)}$ provided 
in Eqs.~(\ref{eq:fnk<}) and (\ref{eq:fnk>}) can be easily rewritten in an easy form
separating charge exchange interactions occurring in the upstream and in 
the downstream. For
instance, let us consider a point upstream of the shock, at $z<0$, and let us
consider first the case $v_{\parallel}<0$. Then, from Eq.~(\ref{eq:fnk<}) one
has:
\begin{eqnarray}
f_{N}^{(k)}(z<0, v_\parallel<0) =&& - \int_{z}^{0}
\frac{dz'}{v_{\parallel}} \beta_{N}^{(k-1)} f_{i}
\exp \left[ \int_{z}^{z'} \frac{dz''}{v_{\parallel}} \beta_{i} \right] -
\int_{0}^{+\infty} \frac{dz'}{v_{\parallel}} \beta_{N}^{(k-1)} f_{i} \exp \left[
\int_{z}^{z'} \frac{dz''}{v_{\parallel}} \beta_{i} \right]  \nonumber \\
\equiv && f_{N,u}^{(k)}+f_{N,d}^{(k)},
\label{eq:broad}
\end{eqnarray}
where $f_{N,u}^{(k)}$ and $f_{N,d}^{(k)}$ are the contributions to
$f_{N}^{(k)}$ deriving from charge exchange events occurred upstream ($u$) and downstream
($d$). 
Similarly, for $v_{\parallel}>0$ and $z<0$, from Eq.~\ref{eq:fnk>}:
\be
f_{N}^{(k)}(z<0,v_\parallel>0) = \int_{-\infty}^{z} \frac{dz'}{v_{\parallel}} \beta_{N}^{(k-1)} 
f_{i} \exp \left[ \int_{z}^{z'} \frac{dz''}{v_{\parallel}} \beta_{i} \right] 
\equiv f_{N,u}^{(k)}.
\ee

In Eq.~(\ref{eq:broad}) one should notice that the fraction of ions in the
distribution function $f_{i}$ with $v_{\parallel}<0$ is tiny, so that to a
good approximation $f_{N}^{(k)}(v_{\parallel}<0,z<0)\approx f_{N,d}^{(k)}$. In
other words the main contribution to the $f_{N}^{(k)}$ with $v_{\parallel}<0$
upstream comes from charge exchange events that have occurred downstream. On the other
hand, $f_{N}^{(k)}(v_{\parallel}>0,z<0)=f_{N,u}^{(k)}$, namely the part of the
distribution function with $v_{\parallel}>0$ is completely determined by the
charge exchange events that occur upstream. Hence, the separation of $f_{N}^{(k)}$ into
the two contributions $f_{N,u}$ and $f_{N,d}$ roughly coincides with the
separation of the distribution function into $f_{N}^{(k)}(z<0,v_{\parallel}<0)$
and $f_{N}^{(k)}(z<0,v_{\parallel}>0)$. A similar line of thought in the
downstream region leads to:
\be
f_{N}^{(k)}(z>0,v_{\parallel}<0) = f_{N,d}^{(k)}
\ee
and
\be
f_{N}^{(k)}(z>0,v_{\parallel}>0) = f_{N,u}^{(k)}+f_{N,d}^{(k)}.
\ee
As stressed in the main part of the paper and illustrated in Fig.~\ref{fig:fn},
the distribution functions are not Maxwellians, neither are Maxwellian the
individual $f_{N}^{(k)}$, although for $k\gg 1$ their shape eventually becomes closer to
that of a Maxwellian. The functions $f_{N,u}^{(k)}$ and
$f_{N,d}^{(k)}$, however, are much more similar to a Maxwellian than the total distribution.
Hence, in order to simplify the calculation of the scattering rates we determine the
moments of $f_{N,u}^{(k)}$ and $f_{N,d}^{(k)}$ and we determine their
contribution to $\beta_{N}^{(k)}$ as the sum of the contributions of Maxwellians
with the same moments as $f_{N,u}^{(k)}$ and $f_{N,d}^{(k)}$. In other words,
for $f_{N,u}^{(k)}$ and $f_{N,d}^{(k)}$ we define:
\be
n_{N,u(d)}^{(k)} = \int d^{3} v f_{N,u(d)}^{(k)},
\ee
\be
v_{N,u(d)}^{(k)}=\frac{1}{n_{N,u(d)}^{(k)}} \int d^{3} v v_{\parallel} 
f_{N,u(d)}^{(k)},
\ee
\be
v_{th,N,u(d)}^{(k)}=\frac{1}{n_{N,u(d)}^{(k)}} \int d^{3} v (v_{\parallel}-
v_{N,u(d)}^{(k)}) f_{N,u(d)}^{(k)}
\ee
and we write:
\be
\beta_{N}^{(k)}(z,v_{\parallel},v_{\perp}) = m_{p} n_{N,u}^{(k)} (z) \sigma_{ce}
(U_{*,N,u}) U_{*,N,u} + m_{p} n_{N,d}^{(k)} (z) \sigma_{ce} (U_{*,N,d})
U_{*,N,d},
\ee
where $U_{*,N,u}=\sqrt{\frac{4}{\pi}{v_{th,N,u}^{(k)}}^{2} + (v_{N,u}^{(k)}-
v_{\parallel})^{2} + v_{\perp}^{2}}$ and
$U_{*,N,d}=\sqrt{\frac{4}{\pi}{v_{th,N,d}^{(k)}}^{2} +
(v_{N,d}^{(k)} - v_{\parallel})^{2} + v_{\perp}^{2}}$.

\end{document}